\documentclass[12pt]{article}
\usepackage[T1]{fontenc}
\usepackage{bm}
\title{How to calculate the main characteristics \\ of random graphs - a new approach}
\author{Agata Fronczak, Piotr Fronczak and Janusz A. Ho\l yst}
\input epsf
\epsfclipon
\input epsf.sty

\begin{document}
\maketitle

\begin{center}
Faculty of Physics and Center of Excellence for Complex Systems
Research, \\ Warsaw University of Technology, \\ Koszykowa 75,
PL-00-662 Warsaw, Poland
\end{center}

\vspace{0.5cm}

\begin{abstract}
The poster presents an analytic formalism describing metric
properties of undirected random graphs with arbitrary degree
distributions and statistically uncorrelated (i.e. randomly
connected) vertices. The formalism allows to calculate the main
network characteristics like: the position of the phase transition
at which a giant component first forms, the mean component size
below the phase transition, the size of the giant component and
the average path length above the phase transition. Although most
of the enumerated properties were previously calculated by means
of generating functions, we think that our derivations are
conceptually much simpler.
\end{abstract}

\begin{center}
{\it A poster presented at Midterm Conference COSIN - \\Conference
on Growing Networks and Graphs\\ in Statistical Physics, Finance,
Biology and Social Systems, \\ Roma 1-5 September 2003.}
\end{center}

\section{Introduction}

Let us start with the following lemma.
\newtheorem{tw}{Lemma}
\begin{tw}\label{tw1}
If $A_{1},A_{2},\dots,A_{n}$ are mutually independent events and
their probabilities fulfill relations $\forall_{i} P(A_{i})\leq
\varepsilon$ then
\begin{equation}
P(\bigcup_{i=1}^{n}A_{i})=1-\exp(-\sum_{i=1}^{n}P(A_{i}))-q,
\end{equation}
where $q<\sum_{j=0}^{n+1} (n\varepsilon)^{j}/j!-(1+\varepsilon)
^{n}$ and may be neglected in the limit of large $n$.
\end{tw}

The complete proof of the Lemma is given in \cite{fronczak}. In
the course of the presentation, we will take advantage of the
Lemma several times.

A random graph with a given degree distribution $P(k)$ is the
simplest network model \cite{newman1}. In such a network the total
number of vertices $N$ is fixed. Degrees of all vertices are
independent, identically distributed random integers drawn from a
specified distribution $P(k)$ and there are no vertex-vertex
correlations. Because of the lack of correlations the probability
that there exists a walk of length $x$ crossing index-linked
vertices $\{i,v_{1},v_{2} \dots v_{(x-1)},j\}$ is described by the
product
$\widetilde{p}_{iv_{1}}\;\widetilde{p}_{v_{1}v_{2}|iv_{1}}\;
\widetilde{p}_{v_{2}v_{3}|v_{1}v_{2}}\dots
\widetilde{p}_{v_{(x-1)}j|v_{(x-2)}v_{(x-1)}}$, where
\begin{equation}\label{pij1}
\widetilde{p}_{ij}=\frac{k_{i}k_{j}}{\langle k\rangle N}
\end{equation}
gives a connection probability  between vertices $i$ and $j$ with
degrees $k_{i}$ and $k_{j}$ respectively, whereas
\begin{equation}\label{pij1b}
\widetilde{p}_{ij|li}=\frac{(k_{i}-1)k_{j}}{\langle k\rangle N}
\end{equation}
describes the conditional probability of a link $\{i,j\}$ given
that there exists {\it another} link $\{l,i\}$. Taking advantage
of the Lemma \ref{tw1} one can write the probability
$p_{ij}^{+}(x)$ of at least one walk of length $x$ between $i$ and
$j$
\begin{eqnarray}\label{pijxA}
p_{ij}^{+}(x)=1-\exp[-\sum_{v_{1}=1}^{N}\sum_{v_{2}=1}^{N}\dots
\sum_{v_{(x-1)}=1}^{N}\widetilde{p}_{iv_{1}} \dots
\widetilde{p}_{v_{(x-1)}j|v_{(x-2)}v_{(x-1)}}].
\end{eqnarray}
Putting (\ref{pij1}) and (\ref{pij1b}) into (\ref{pijxA}) and
replacing the summing over nodes indexes by the summing over the
degree distribution $P(k)$ one gets:
\begin{equation}\label{pijxB}
p_{ij}^{+}(x)=1-\exp\left[-\frac{k_{i}k_{j}}{N}\frac{\langle
k(k-1) \rangle^{x-1}}{\langle k\rangle^{x}}\right].
\end{equation}

\section{Random graphs below the percolation \\
threshold - the mean component size}

According to (\ref{pijxB}), the probability that none among the
walks of length $x$ between $i$ and $j$ occurs is given by
\begin{equation}\label{pijxC}
p_{ij}^{-}(x)=1-p_{ij}^{+}(x)=\exp\left[-\frac{k_{i}k_{j}}{N}\frac{\langle
k(k-1) \rangle^{x-1}}{\langle k\rangle^{x}}\right]
\end{equation}
and respectively the probability that there is no walk of any
length between these vertices may be written as
\begin{eqnarray}\label{pijm1}
p_{ij}^{-}=\prod_{x=1}^{\infty}\;p_{ij}^{-}(x)=
\prod_{x=1}^{\infty}\;\exp\left[-\frac{k_{i}k_{j}}{N}\frac{\langle
k(k-1) \rangle^{x-1}}{\langle k\rangle^{x}}\right] \;= \;\nonumber
\\=\;\exp\left[-\;\frac{k_{i}k_{j}}{\langle k\rangle N}\;
\sum_{y=0}^{\infty}\;\left(\frac{\langle k^{2}\rangle}{\langle k
\rangle}-1\right)^{y}\right].
\end{eqnarray}
The value of $\;p_{ij}^{-}\;$ strongly depends on the common ratio
of the geometric series present in the last equation. When the
common ratio is greater then $1$ i.e. $\langle k^{2}\rangle\geq
2\langle k\rangle$ random graphs are above the percolation
threshold. The sum of the geometric series in (\ref{pijm1}) tends
to infinity and $p_{ij}^{-}=0$. Below the phase transition, when
$\langle k^{2}\rangle<2\langle k\rangle$, the probability that the
nodes $i$ and $j$ belong to separate clusters is given by
\begin{equation}\label{pijm}
p_{ij}^{-}\;=\;\exp\left[-\;\frac{k_{i}k_{j}}{N}\frac{1}{(2\langle
k\rangle-\langle k^{2}\rangle)}\right]
\end{equation}
and respectively the probability that $i$ and $j$ belong to the
same cluster may be written as
\begin{equation}\label{pijp}
p_{ij}^{+}\;=\;1-p_{ij}^{-}\;=\;1-\exp\left[-\;\frac{k_{i}k_{j}}{N}\frac{1}{(2\langle
k\rangle-\langle k^{2}\rangle)}\right].
\end{equation}

\begin{figure} \epsfxsize=9 cm \epsfbox{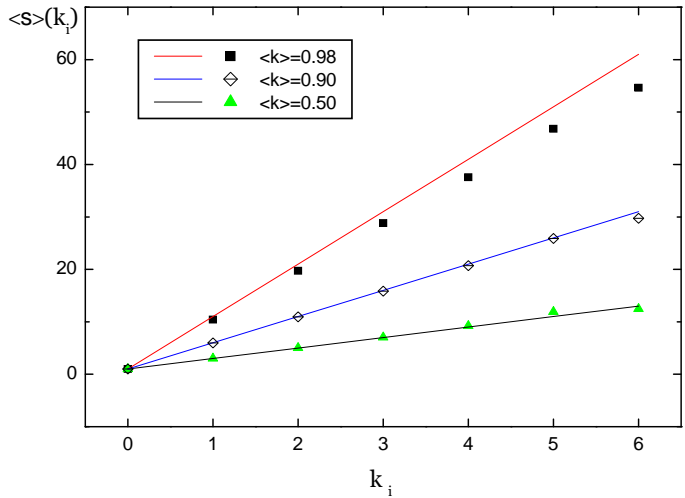}
\caption{Average size of the component that a node $i$ with degree
$k_{i}$ belongs to. Scatter plots represent numerical data,
whereas solid lines represent the prediction of Eq. (\ref{skk}).}
\label{figERskk}
\end{figure}

Now, it is simple to calculate the mean size of the cluster that
the node $i$ belongs to. It is given by
\begin{equation}\label{skk}
\langle s \rangle(k_{i})\;=\;\sum_{k_{j}}P(k_{j})p_{ij}^{+}+1
\;\simeq\; \frac{\langle k\rangle}{2\langle k\rangle-\langle
k^{2}\rangle}k_{i}+1.
\end{equation}
Note, that the mean size of the component that a node $i$ belongs
to, is proportional to degree $k_{i}$ of the node (see Fig.
\ref{figERskk}). The last transformation in (\ref{skk}) was
obtained by taking only the first two terms of power series
expansion of the exponential function in (\ref{pijp}). Averaging
the above expression (\ref{skk}) over all nodes in the network one
obtains the well-known formula \cite{newman1} for the mean
component size in random graphs below the phase transition
\begin{equation}\label{sk}
\langle s \rangle \;=\; 1+\frac{\langle k\rangle^{2}}{2\langle
k\rangle-\langle k^{2}\rangle}.
\end{equation}
As in percolation theory \cite{stauffer}, the mean cluster size
diverges at
\begin{equation}\label{pc}
\langle k^{2}\rangle=2\langle k\rangle,
\end{equation}
signifying that the expression (\ref{pc}) describes the position
of the percolation threshold in random graphs with arbitrary
degree distributions \cite{newman1,molloy1,cohen_pc}.

\section{Random graphs above the percolation \\ threshold -
the size of the giant component}

When $\langle k^{2}\rangle>2\langle k\rangle$ the giant component
(GC) is present in the graph. The size of the giant component
$N_{GC}$ scales as the size of the graph as a whole $N$. Its
relative size $S=N_{GC}/N$ (i.e. the probability that a node
belongs to GC) is an important quantity in percolation theory and
is often identified as the order parameter. Here we demonstrate
how to calculate the size of the giant component in undirected
random graphs. The underlying concept, how to calculate $S$, is
closely related to the method of calculating $S$ in Cayley tree
presented in \cite{stauffer}.

At the beginning, we deal with classical random graphs of
Erd\"{o}s and R\'{e}nyi, then we generalize our derivations for
the case of random graphs with arbitrary degree distributions and
we show that our derivations are consistent with the formalism
based on generating functions that was introduced by Newman {\it
et al.} \cite{newman1}.

\subsection{The giant component in classical random graphs \\ of Erd\"{o}s and R\'{e}nyi}

In general terms, classical random graphs consist of a fixed
number of vertices $N$, which are connected at random with a fixed
probability $p$ \cite{bollobas}.

Let us call $R$ the probability that an arbitrary node $i$ is
connected to the giant component through a fixed link $\{i,j\}$,
where $j$ is another arbitrary node. Since every node in the graph
may have $N-1\simeq N$ links and all nodes are equivalent, the
formula for $R$ may be written as the product of the probability
of a link $\{i,j\}$ and the probability that at least one of $N$
possible links emanating from $j$ connects $j$ to the giant
component. Taking advantage of the Lemma \ref{tw1} one can write
\begin{equation}\label{Rer}
R\;=\;p\;(1-\exp[-RN]).
\end{equation}

\begin{figure} \epsfxsize=9 cm \epsfbox{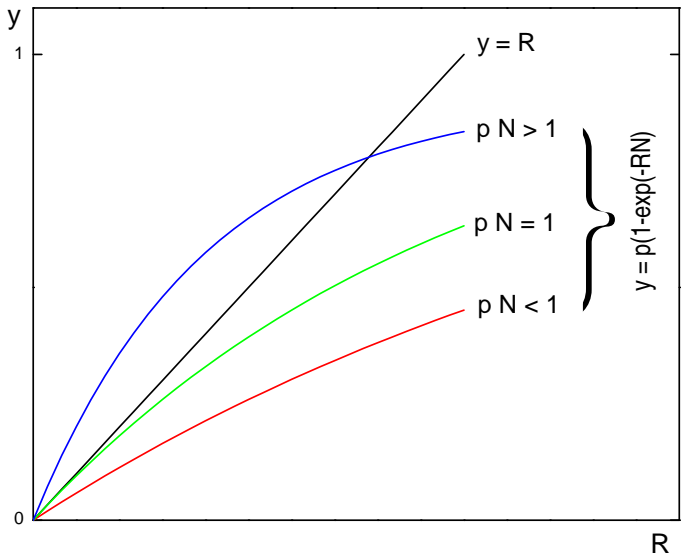}
\caption{Graphical solution of the equation (\ref{Rer}).}
\label{figERpc}
\end{figure}

This self-consistency equation for $R$ has one or two solutions,
depending on whether a graph is below ($pN<1$) or above ($pN>1$)
the phase transition. Graphical solution of the equation
(\ref{Rer}) shown at Fig. \ref{figERpc} presents the easiest way
to obtain a qualitative understanding of percolation transition in
classical random graphs.

\begin{figure} \epsfxsize=9 cm \epsfbox{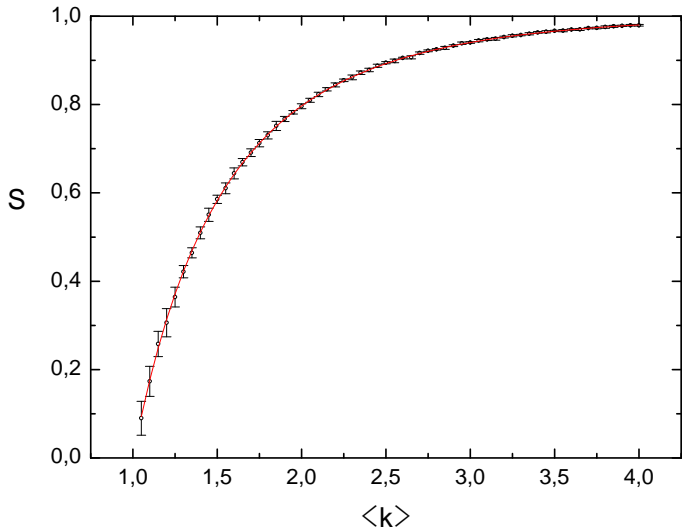}
\caption{The size of the giant component $S$ versus the mean node
degree $\langle k\rangle$ in classical random graphs of size
$N=10000$. The scatter plot represents numerical data whereas the
solid line gives the solution of the Eq.(\ref{Ser2})}.
\label{figERS}
\end{figure}

The probability that an arbitrary node $i$ belongs to the giant
component is equivalent to the probability that at least one of
$N$ possible links connects $i$ to GC. Again, taking advantage of
the Lemma 1. one gets
\begin{equation}\label{Ser1}
S\;=\;1-\exp[-RN].
\end{equation}
Comparing both relations (\ref{Rer}) and (\ref{Ser1}) it is easy
to see that $R=pS$ and the expression (\ref{Ser1}) for the giant
component in classical random graphs may be rewritten in the form
\begin{equation}\label{Ser2}
S\;=\;1-\exp[-\langle k\rangle S],
\end{equation}
where $\langle k\rangle=pN$. Fig. \ref{figERS} presents the
prediction of the Eq. (\ref{Ser2}) in comparison with numerically
calculated sizes of the giant components in classical random
graphs.

It is necessary to stress that both equations (\ref{Rer}) and
(\ref{Ser1}) are well-known and have been independently derived
using different methods by Newman {\it et al.} \cite{newman1} and
Molloy and Reed \cite{molloy2}.

\subsection{The giant component in random graphs \\
with arbitrary degree distributions $P(k)$}

In the case of classical random graphs all vertices have been
considered as equivalent. It is not acceptable in the case of
random graphs with a given degree distribution $P(k)$, where each
node $i$ is characterized by its degree $k_{i}$.

Here, we call $R^{*}$ the probability that following arbitrary
direction of a randomly chosen edge we arrive at the giant
component. In fact, we know that following an arbitrary edge we
arrive at a vertex $i$ with degree $k_{i}$. The probability that
$i$ is connected to GC is $1-(1-R^{*})^{k_{i}-1}$. The notation
expresses the probability that at least one of $k_{i}-1$ edges
emanating from $i$ and other than the edge we arrived along
connects $i$ to the giant component\footnote{We do not here take
advantage of the Lemma \ref{tw1} because of it works well the
limit of large number of independent events $n\gg 1$. In the case
of small $n$ the error $q$ of the Lemma \ref{tw1} can not be
neglected.}. Now, it is simple to write the self-consistency
condition for $R^{*}$
\begin{equation}\label{Rrg}
R^{*}\;=\;\sum_{k_{i}}\left(1-(1-R^{*})^{k_{i}-1}\right)Q(k_{i}),
\end{equation}
where $Q(k_{i})=k_{i}P(k_{i})/\langle k\rangle$ describes the
probability that an arbitrary link leads to a node $i$ with degree
$k_{i}$. As in the case of classical random graphs the equation
(\ref{Rrg}) can be solved graphically, signifying that the
nontrivial solution (i.e. $R^{*}\neq 0$) of the equation
(\ref{Rrg}) exists only for random graphs above the percolation
threshold $\langle k^{2}\rangle>2\langle k\rangle$.

Knowing $R^{*}$, it is simple to calculate the relative size $S$
of the giant component in random graphs with arbitrary degree
distribution $P(k)$. $S$ is equivalent to the probability that at
least one of $k$ links attached to an arbitrary node connects the
node to the giant component
\begin{equation}\label{Srg}
S\;=\;\sum_{k}\left(1-(1-R^{*})^{k}\right)P(k).
\end{equation}

It is easy to show that both equations (\ref{Srg}) and (\ref{Rrg})
are completely equivalent to equations derived by Newman {\it et
al.} by means of generating functions
\begin{equation}\label{new1}
S=1-G_{0}(v),
\end{equation}
where $v$ is the solution of equation given below
\begin{equation}\label{new2}
v=G_{1}(v).
\end{equation}
We recall that $G_{0}(x)$ is the generating function for the
probability distribution of vertex degrees
\begin{equation}\label{G0}
G_{0}(x)=\sum_{k}P(k)x^{k},
\end{equation}
whereas $G_{1}(x)$ is given by
\begin{equation}\label{G1}
G_{1}(x)=\frac{G^{'}_{0}(x)}{\langle k\rangle}= \frac{1}{\langle
k\rangle}\sum_{k}kP(k)x^{k-1}.
\end{equation}

At the beginning we show that Eq. (\ref{Rrg}) is completely
equivalent to Eq. (\ref{new2}). Expression (\ref{Rrg}) may be
transformed in the following way
\begin{eqnarray*}
R^{*} = \frac{1}{\langle k \rangle}\sum_{k}k P(k) -
\frac{1}{\langle k\rangle}\sum_{k}k P(k) (1-R^{*})^{k-1}
=1-G_{1}(1-R^{*}),
\end{eqnarray*}
that exactly corresponds to Eq. (\ref{new2}) with $v=1-R^{*}$.
Expression (\ref{Srg}) may be transformed into Eq. (\ref{new1}) in
a similar way, when assume that $v=1-R^{*}$. Now, it is clear that
the unknown parameter $v$ in both Eqs. (\ref{new1}) and
(\ref{new2}) has the following meaning:
\begin{equation}\label{vR}
v=1-R^{*}
\end{equation}
describes the probability that an arbitrary edge in random graph
does not belong to the giant component.

\section{Average path length in random graphs}

This part of the presentation closely follows that of Fronczak
{\it et al.} \cite{fronczak}.

Let us consider the situation when there exists at least one walk
of the length $x$ between the vertices $i$ and $j$. If the walk(s)
is(are) the shortest path(s) $i$ and $j$ are exactly $x$-th
neighbors otherwise they are closer neighbors. In terms of
statistical ensemble of random graphs the probability
$p_{ij}^{+}(x)$ (Eq. (\ref{pijxB})) of at least one walk of the
length $x$ between $i$ and $j$ expresses also the probability that
these nodes are neighbors of order not higher than $x$. Thus, the
probability that $i$ and $j$ are exactly $x$-th neighbors is given
by the difference\footnote{Note, that (\ref{pijx*A}) is only true
for random graphs above the percolation threshold where
$p_{ij}^{*}(x)>0$.}
\begin{equation}\label{pijx*A}
p_{ij}^{*}(x)=p_{ij}^{+}(x)-p_{ij}^{+}(x-1).
\end{equation}

Due to (\ref{pijxB}) the probability that both vertices are
exactly the $x$-th neighbors may be written as
\begin{equation}\label{pijx*B}
p_{ij}^{*}(x)=F(x-1)-F(x),
\end{equation}
where
\begin{equation}\label{Fx}
F(x)=\exp \left[ -\frac{k_{i}k_{j}}{N} \frac{(\langle
k^{2}\rangle-\langle k\rangle)^{x-1}}{\langle k\rangle^{x}}
\right].
\end{equation}
Now, it is simple to calculate the average path length (APL)
between $i$ and $j$. It is given by
\begin{equation}\label{lijA}
l_{ij}(k_{i},k_{j})=\sum_{x=1}^{\infty}\:x\:p_{ij}^{*}(x)=\sum_{x=0}^{\infty}F(x).
\end{equation}
Notice that a walk may cross the same node several times thus the
largest possible walk length can be $x=\infty$.

The Poisson summation formula
\begin{equation}\label{poissonform}
\sum_{x=0}^{\infty}F(x)=\frac{1}{2}F(0)+\int_{0}^{\infty}F(x)dx
+2\sum_{n=1}^{\infty}\left(\int_{0}^{\infty}F(x)\cos(2n\pi
x)dx\right)
\end{equation}
allows us to simplify (\ref{lijA}). Firstly, let us note that in
most of real networks $\langle k\rangle\ll N$ thus we can assume
\begin{equation}\label{assumption}
\frac{k_{i}k_{j}}{N\langle k^{2}\rangle}\simeq 0
\end{equation}
that gives $F(0)=1$. Secondly, we have
\begin{equation}\label{Ei}
\int_{0}^{\infty} F(x)dx = -Ei\left(-\frac{k_{i}k_{j}}{\langle
k^{2}\rangle N}\right)/\ln\left(\frac{\langle
k^{2}\rangle}{\langle k\rangle}\right),
\end{equation}
where $Ei(y)$ is the exponential integral function that for
negative arguments is given by $Ei(-y)=\gamma+\ln y +
\int_{0}^{y}(\exp(-t)-1)/t\:dt$ \cite{40}, where $\gamma\simeq
0.5772$ is the Euler's constant. Due to (\ref{assumption}) the
integral in the expression for $Ei(-y)$ becomes zero. Finally,
every integral in the last term of the summation formula
(\ref{poissonform}) is equal to zero owing to the generalized mean
value theorem \cite{43}. It follows that the equation for the APL
between $i$ and $j$ may be written as
\begin{equation}\label{lij}
l_{ij}(k_{i},k_{j})=\frac{-\ln k_{i}k_{j}+\ln(\langle
k^{2}\rangle-\langle k\rangle)+\ln N -\gamma}{\ln(\langle
k^{2}\rangle/\langle k\rangle-1)}+\frac{1}{2}.
\end{equation}

The average intervertex distance for the whole network depends on
a specified degree distribution $P(k)$
\begin{equation}\label{lRG}
l=\frac{\ln(\langle k^{2}\rangle-\langle k\rangle)-2\langle\ln
k\rangle+\ln N -\gamma}{\ln(\langle k^{2}\rangle/\langle
k\rangle-1)}+\frac{1}{2}.
\end{equation}
A similar result $l\sim \ln N/\ln(\langle k^{2}\rangle/\langle k
\rangle-1)$ was obtained by Dorogovtsev {\it et al.}
\cite{metric}. The formulas (\ref{lij}) and (\ref{lRG}) diverge
when $\langle k^{2}\rangle=2\langle k\rangle$, giving the
well-known expression for percolation threshold in undirected
random graphs (\ref{pc}).

\subsection{Average path length in classical random graphs \\ of
Erd\"{o}s and R\'{e}nyi}

For these networks the degree distribution is given by the Poisson
function $P(k)=e^{-\langle k\rangle}\langle k\rangle^{k}/k!$.
However, since $\langle\ln k\rangle$ cannot be calculated
analytically for Poisson distribution thus the $APL$ may not be
directly obtained from (\ref{lRG}). To overcome this problem we
take advantage of the mean field approximation. Let us assume that
all vertices within a graph possess the same degree
$\forall_{i}\:k_{i}=\langle k\rangle$. It implies that the $APL$
between two arbitrary nodes $i$ and $j$ (\ref{lRG}) should
describe the average intervertex distance of the whole network
\begin{equation}\label{lER}
l_{ER}=\frac{\ln N - \gamma}{\ln(pN)}+\frac{1}{2}.
\end{equation}

\begin{figure} \epsfxsize=9 cm \epsfbox{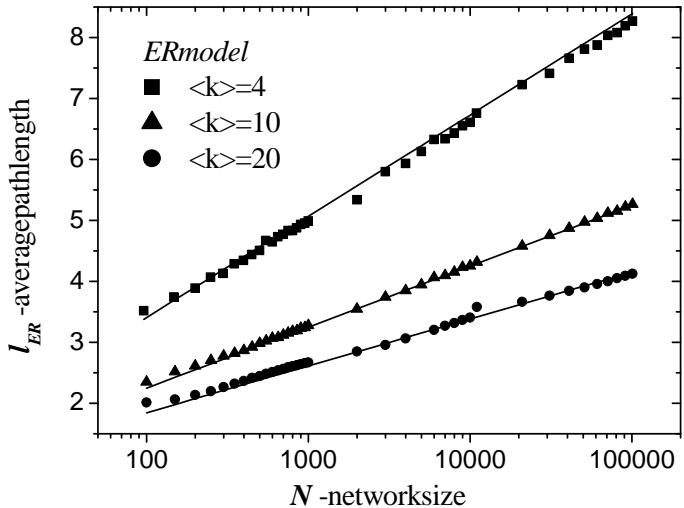}
\caption{The average path length $l_{ER}$ versus network size $N$
in $ER$ classical random graphs with $\langle k \rangle
=pN=4,10,20$. Solid curves represent numerical prediction of
Eq.(\ref{lER}).} \label{figer}
\end{figure}

Until now only a rough estimation of the quantity has been known.
One has expected that the average shortest path length of the
whole ER graph scales with the number of nodes in the same way as
the network diameter. We remind that the diameter $d$ of a graph
is defined as the maximal distance between any pair of vertices
and  $d_{ER}=\ln N/\ln(pN)$. Fig.\ref{figer} shows the prediction
of the equation (\ref{lER}) in comparison with the numerically
calculated $APL$ in classical random graphs.

\subsection{Average path length in scale-free \\ Barab\'{a}si-Albert
networks}

The basis of the $BA$ model is its construction procedure. Two
important ingredients of the procedure are: continuous network
growth and preferential attachment. The network starts to grow
from an initial cluster of $m$ fully connected vertices. Each new
node that is added to the network creates $m$ links that connect
it to previously added nodes. The preferential attachment means
that the probability of a new link growing out of a vertex $i$ and
ending up in a vertex $j$ is given by
$\widetilde{p}_{ij}^{BA}=mk_{j}(t_{i})/\sum_{l}k_{l}(t_{i})$,
where $k_{j}(t_{i})$ \cite{42} denotes the connectivity of a node
$j$ at the time when a new node $i$ is added to the network.
Taking into account the time evolution of node degrees in $BA$
networks one can show  that the probability
$\widetilde{p}_{ij}^{BA}$ is equivalent to (\ref{pij1}). Now let
us consider the conditional probability $\widetilde{p}_{ij|li}$.
Checking the possible time order of the vertices $i,j,l$ it is
easy to see that in five of $3!$ cases
$\widetilde{p}_{ij|li}=\widetilde{p}_{ij}$ and in a good
approximation we get instead of (\ref{pijxB}) the result
\begin{equation}\label{pijxBA}
p_{ij}^{+}(x)=1-\exp \left[ -\frac{k_{i}k_{j}}{N} \frac{\langle
k^{2}\rangle^{x-1}}{\langle k\rangle^{x}} \right].
\end{equation}

\begin{figure} \epsfxsize=9 cm \epsfbox{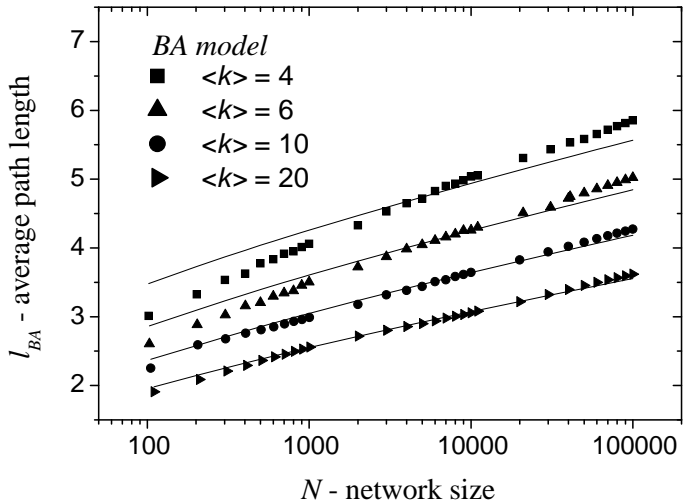}
\caption{Characteristic path length $l_{BA}$ versus network size
$N$ in $BA$ networks. Solid lines represent Eq.(\ref{lBA}).}
\label{figba}
\end{figure}

It was found \cite{42} that the degree distribution in $BA$
network is given by $P(k)=2m^{2}k^{-\alpha}$, where
$k=m,m+1,\dots,m\sqrt{N}$, and the scaling exponent $\alpha=3$.
Putting $\langle k\rangle=2m$, $\langle k^{2}\rangle=m^{2}\ln N$
and taking into account (\ref{pijxBA}) one gets that the $APL$
between $i$ and $j$ is given by
\begin{equation}\label{lijBA}
l_{ij}^{BA}(k_{i},k_{j})=\frac{-\ln(k_{i}k_{j})+\ln N+\ln(2m)
-\gamma }{\ln\ln N+\ln (m/2)}+\frac{3}{2}.
\end{equation}
Averaging (\ref{lijBA}) over all vertices we obtain
\begin{equation}\label{lBA}
l_{BA}=\frac{\ln N-\ln(m/2)-1-\gamma}{\ln\ln
N+\ln(m/2)}+\frac{3}{2}.
\end{equation}
Fig.\ref{figba} shows the $APL$ of $BA$ networks as a function of
the network size $N$ compared with the analytical formula
(\ref{lBA}). There is a visible discrepancy between the theory and
numerical results when $\langle k\rangle=4$. The discrepancy
disappears when the network becomes denser i.e. when $\langle
k\rangle$ increases.

\subsection{Average path length in scale-free networks \\ with arbitrary
scaling exponent}

Let us consider scale-free random graphs with degree distribution
given by a power law, i.e.
$P_\alpha(k)=(\alpha-1)m^{\alpha-1}k^{-\alpha}$, where
$k=m,m+1,\dots,mN^{1/(\alpha-1)}$ \cite{9}. Taking advantage of
(\ref{lRG}) we get that for large networks $N\gg 1$ the $APL$
scales as follows
\begin{itemize}
\item $l\simeq 2/(3-\alpha)+1/2\:\:\:$ for $2<\alpha<3$,
\item $l\simeq \ln N/\ln\ln N+3/2\:\:\:$ for $\alpha=3$,
\item $l\simeq \ln N/(\ln(m(\alpha-2)/(\alpha-3)-1)+1/2\:\:\:$ for $\alpha>3$.
\end{itemize}
The result for $\alpha\geq 3$ is consistent with estimations
obtained by Cohen and Havlin \cite{9}. The first case with $l$
independent on $N$ shows that there is a saturation effect for the
mean path length in large scale-free networks with scaling
exponent from the range $2<\alpha<3$. Our derivations show that
the behaviour of $APL$ within scale-free networks is even more
intriguring than reported by Cohen and Havlin \cite{9}.


\end{document}